\documentclass{acm_sen_article}
\usepackage{cite}
\usepackage{graphicx}
\usepackage{amssymb, amsmath}
\usepackage{balance}
\usepackage{hyperref}
\hypersetup{
  hidelinks,
}
\begin{document}

\title{Reflections on the Evolution of Computer Science Education}

\numberofauthors{1} 
\author{
Sreekrishnan Venkateswaran\\
       \affaddr{Kyndryl Corporation}\\
       \affaddr{Bangalore, India}\\
       \email{s.krishna@kyndryl.com}
}
\maketitle

\begin{abstract}
Computer Science education has been evolving over the years to reflect applied realities. Until about a decade ago, theory of computation, algorithm design and system software dominated the curricula. Most courses were considered core and were hence mandatory; the programme structure did not allow much of a choice or variety. This column analyses why this changed Circa 2010 when elective subjects across scores of topics become part of mainstream education to reflect the on-going lateral acceleration of Computer Science.

Fundamental discoveries in artificial intelligence, machine learning, virtualization and cloud computing are several decades old. Many core theories in data science are centuries old. Yet their leverage exploded only after Circa 2010, when the stage got set for people-centric problem solving in massive scale. This was due in part to the rush of innovative real-world applications that reached the common man through the ubiquitous smart phone. AI/ML modules arrived in popular programming languages; they could be used to build and train models on powerful - yet affordable - compute on public clouds reachable through high-speed Internet connectivity. Academia responded by adapting Computer Science curricula to align it with the changing technology landscape.

The goal of this experiential piece is to trigger a lively discussion on the past and future of Computer Science education.

\end{abstract}

\section{The Evolution}

When I was a student over a quarter of a century ago, Computer Science (CS) education was about learning finite automata, intractability, formal language theory, compilers and assemblers. They taught us the Turing Test, the Halting Problem and Russell’s Paradox. We recursively negotiated the Eight Queens Puzzle and grappled with the Traveling Salesman Problem~\cite{Cormen} in our programming labs. We were asked to design compilers for hypothetical programming languages for our semester projects. Lazy evaluations and Greedy approaches were well-known algorithmic paradigms; no one would misunderstand them to be human behavioural patterns! Getting industry-ready for a CS student meant acquiring the ability to develop and debug Operating System (OS) code and design look-ahead parsers. Campus interview questions would entail designing fast-performing system software algorithms or a scholarly comparison of programming languages.

During my Master’s in CS, when our department purchased new DEC Alpha\footnote{Alpha is the name of the RISC server developed in the 1990s by Digital Equipment Corporation (DEC)} systems for the computer lab, we had to cross-compile the compiler before we could start indulging on our programming experiments. We were only starting to figure out what the Internet was, so we neither had to attend courses on social network analysis nor master the Web. And because the online learning economy would not be born for another 15 years, we had to pay attention to what our professors were expounding in class.

When I joined my first job from the campus, I found high-quality alignment between what we had been taught and the customer projects that we were assigned to. We had to, for example, contend with the Dining Philosopher’s Problem\footnote{An example scenario developed by Edsger Dijkstra to demonstrate how to design concurrent algorithms} when doing kernel programming, and apply graph theory to optimally traverse network routing tables. We did ‘C’ programming to code Linux device drivers, developed firmware for OS-less devices in assembly language, and wrote VHDL\footnote{Very high-speed integrated circuit Hardware Description Language (VHDL) is used to specify the logic design of a configurable digital system such as a Complex Programmable Logic Device (CPLD)} code to optimize software partitioning between the hardware and the OS. My life revolved around kernel debuggers, logic analysers and oscilloscopes as we did software-hardware co-design for embedded systems. The company’s University Relations program leveraged campus industry labs to test our network protocol code for interoperability with systems built by other vendors.

\noindent \textbf{Circa 2010:} Sands started to shift. Software began to touch every facet of life. Application development turned mainstream. The demand for designing complex user-centric software started outpacing requirements for engineering system software. API\footnote{Application Programming Interface} Economy~\cite{api-economy} took root as a technology design point as the world economy digitized across industries and sectors. Cloud computing, as-a-service consumption, and pay-as-you-go billing started changing information technology. The demand for OS internals programming, deeply embedded firmware, and system software reduced and they gravitated to the niche.

Artificial Intelligence (AI) and Machine Learning (ML) had been in existence for decades. Neural Networks were first proposed in the 1940s, virtualization began in the 1960s and early forms of cloud computing came soon after. Core discoveries associated with statistical models and distributions were even older. The binomial theorem dated back several centuries before it was generalized in the 1700s; the Central Limit Theorem (CLT) and Bayes' Theorem had also been known since the 1700s. The Poisson distribution had been published in 1837; the first Logistic Regression model had also been developed during this period.

But the time had finally arrived for their adoption to explode, thanks to the new-found ease of consumption. Most popular programming languages implemented AI/ML libraries, and they could be trained and delivered through affordable processing power on the cloud reachable through high-speed Internet connectivity. Smart phones brought a profusion of AI/ML use cases to real-world applications accessed by the common man.

People-centric problem solving needs, thus, far surpassed system-centric demands. Scalability requirements hence expanded exponentially. Abstraction for easy use and reuse of software also became maximalist. A programmer needed to only know API specifications in order to leverage the technology in question. Software systems became integration-friendly by design, opening up new possibilities.

Platform-as-a-Service (PaaS) and library modules came of age. They masked the theory and the design of complex algorithms, encapsulated the intrinsic, and enveloped it with an interface engineered for easy use and portability. You no longer designed code to concurrently search billions of distributed records in the blink of an eye; instead, you used the best-available-fit existing service as your starting point. You did not have to write an ML model when you needed to automatically track moving objects from a live video feed in real-time; rather, you chose from existing models that your favourite programming language supported as importable modules, trained the chosen model with datasets that you could find or buy, and tuned hyperparameters until you reached acceptable performance levels.

Software that can generate software gave rise to the low-code/no-code movement that sought to enable application development with minimal coding. A user could visually produce certain types of applications simply by dragging and dropping requirements or components into a graphical interface.

\section{Ensuing Shifts in CS Curricula}

All these signalled a paradigm shift in CS education. Universities started adjusting curricula to minimize the ensuing angle of deviation between academics and industry. 

Elective courses were scarce in the CS curricula of the 1990s. When I did my Bachelors in CS during that period, we had $21$ mandatory core CS courses supplemented by $2$ electives. All electives were department/disciplinary; the concept of open (or inter-departmental) electives did not exist. We could choose the $2$ electives out of a list of $3$: Image Processing, Computer Graphics and Computer Architecture. Today, the same CS program~\cite{nitc-cs} has restructured to require $14$ mandatory core CS courses along with $9$ electives. The electives can be chosen from a large basket of intra and inter departmental courses that include Natural Language Processing, Web Programming, Cloud Computing, Data Mining, Machine Learning, Medical Imaging, Hashing Techniques for Big Data, DNA Computing, Bioinformics, Computer Vision and Embedded Systems. 

Most electives specify a set of other courses as prerequisites, which introduces dependencies by design. Only certain permutations of courses can thus be chosen, hence only a predetermined set of programme outcomes are possible. If you consider courses as nodes, prerequisites generate edges between them and create a directed graph; only a preplanned set of end nodes (or specializations) exist.

This evolution in structure and content largely holds in CS schools across the world~\cite{NUS-cs}\cite{iitk-cs}\cite{stanford-cs}. The end-result is the alignment of university education with the recent lateral expansion of Computer Science.

Let us zoom in on Software Engineering. Two decades back, Software Engineering was rarely part of the CS curriculum. My institute did not offer it. But today, it is often a mandatory part of CS curricula world-wide. The course content, by default, covers Agile frameworks and iterative development. More modern courses on software engineering include the twelve-factor methodology~\cite{12factor} for building software-as-a-service applications and the microservices architecture. They bring out the new world model that expounds the \emph{deliver-fail-change-succeed} cycle where the way to a successful product is to bring out some features, test the market in a calibrated manner that limits the impact radius of failure, and expand iteratively.

Massive Open Online Courses (MOOC) further democratized software skilling and demystified AI usage, though not the core intrinsic of the technology.

\section{Business Alignment}

Simultaneously, a new dimension entered CS education: the notion of Return on Investment (RoI). Business was no longer the bounded domain of management graduates. Engineers were now expected to innovate with cost-awareness because of two reasons. First, the pressure from stock markets in the new-age economy and the explosion of start-up ecosystems seeking to quickly monetize technology became considerable. Delivering in real-time and performing longer-term research now had to progress at a ratio that would keep investors happy. Second, with the ease of as-you-go consumption of the cloud and the API Economy, it became easy for spending to exceed budgets, and this called for cost-conscious software engineering.

Moreover, digitization started touching and changing the core of business models. The classic example is how App-based ridesharing changed the business model of the cab industry with cloud, AI and analytics. Disruption at the business model level extended across industries, from finance and manufacturing to telecom and retail.

\section{Is Core CS Research Diminishing?}
This is not to suggest that core CS research is diminishing in real terms. It is growing indeed. It is not uncommon to see a blockchain engineer model failure-safety after say, the Byzantine Generals Problem~\cite{byz-lamport}. Or to listen to a distributed applications programmer worry about the consistency and durability of a financial transaction.
A chatbot developer has to indeed dwell on modelling ACID\footnote{Atomicity - Consistency - Isolation - Durability are commonly abbreviated as ACID} in order to ensure data integrity while supporting massive concurrency. Some applications have to be designed to operate in near planet-scale, which may require a prowess in algorithm design.

The low-code/no-code movement might have taken off, but it still needs Computer Scientists to design and write programs that write programs! 

In percentage terms, however, the quantum of user-centric engineering and up-the-stack application programming seem to have decisively overtaken core CS and system software development functions today.

\section{Concluding Thoughts}

Any 35-year career cycle is unlikely to be monotonic. Some fluctuations and at least a couple of disruptions are near-certain. However, basic skills usually endure. Whether you code in Python or JavaScript or C, you can connect your work to principles of programming languages if you look one level deeper. Whether you explore a new ML model or perform processing around an existing model to solve a real-world problem, Donald Knuth’s Art of Computer Programming~\cite{knuth} is likely to be relevant. Dijkstra’s shortest path algorithm~\cite{dijkstra} is as relevant to Google Maps and to help chart flight paths as it has been to network routing.

Charles Darwin theorized that it is not the strongest species that survive, but the most adaptable ones. A combination of mastery of basic skills and the propensity to adapt and transform will help not just survive a long career in technology, but to thrive and flourish!

\balance

\bibliographystyle{plain}
\bibliography{cs}

\end{document}